\documentstyle[12pt]{article}  
\def\lsim{\mathrel{\lower2.5pt\vbox{\lineskip=0pt\baselineskip=0pt
          \hbox{$<$}\hbox{$\sim$}}}}
\def\gsim{\mathrel{\lower2.5pt\vbox{\lineskip=0pt\baselineskip=0pt
          \hbox{$>$}\hbox{$\sim$}}}}

\begin{document}   

\markright{C. Barcel\'o, Wormholes in spacetimes with cosmological 
horizons.}

\begin{center}
{\Large\bf Wormholes in spacetimes with cosmological horizons}
\\[3mm]
{Carlos Barcel\'o\\[2mm]
{\small \it Instituto de Astrof\'{\i}sica de 
Andaluc\'{\i}a, CSIC\\
Camino Bajo de Hu\'etor, 18080 Granada, Spain}\\[2mm]
{\small \it and}\\[2mm]
{\small \it Instituto de Matem\'aticas y 
F\'{\i}sica Fundamental, CSIC\\
C/  Serrano 121, 28006 Madrid, Spain}}\\[2mm]
{\small 9 February 1998}\\[5mm]
\end{center}

\begin{abstract}

A generalisation of the asymptotic wormhole boundary condition for 
the case of spacetimes with a cosmological horizon is proposed.
In particular, we consider de Sitter spacetime with small 
cosmological constant. The wave functions selected by this 
proposal are exponentially damped in WKB 
approximation when the scale factor is large but still much smaller
than the horizon size. In addition, they only include outgoing 
gravitational modes in the region beyond the horizon. We argue that 
these wave functions represent quantum wormholes and compute the 
local effective interactions induced by them in low-energy field 
theory. These effective interactions differ from those for flat 
spacetime in terms that explicitly depend on the cosmological 
constant.

\vspace*{5mm}
%Keywords: wormholes, de Sitter horizon, effective interactions
\end{abstract}

\section{Introduction}
\setcounter{equation}{0}
\label{i}

Wormholes are spacetime fluctuations that involve baby universes
branching off and joining onto different regions of spacetime. In 
the dilute wormhole approximation, each wormhole end is considered 
to be connected to a different asymptotically large region of 
spacetime \cite{col1,haw}. In this situation, wormholes can be 
represented quantum mechanically by wave functions that satisfy the 
Wheeler-De Witt (WDW) equation. In order to recover the semiclassical
behaviour expected for a wormhole, Hawking and Page \cite{pag} 
proposed that the wormhole wave functions should be exponentially 
damped for large three-geometries. Besides, these wave functions 
should be regular when the three-geometry degenerates to zero 
\cite{pag}. These boundary conditions are usually employed to select
the wormhole wave functions among the solutions to the WDW equation.

It has been claimed that the existence of wormhole insertions in 
spacetime introduces local effective interactions in low-energy field
theory and may modify the constants of nature \cite{col1,haw1}. 
The analysis of the wormhole effects in spacetimes with cosmological
horizons is particularly relevant. Spacetimes that 
describe solutions of interest in cosmology usually posses this kind 
of horizons. In addition, these horizons are generally present when 
the cosmological constant, $\Lambda$, is positive. Actually, the 
existence of quantum wormholes in spacetimes with 
positive $\Lambda$ was already assumed by Coleman when putting 
forward his mechanism for the vanishing of the observed cosmological 
constant \cite{col2}. However, when one tries to study wormhole 
processes in spacetimes with a cosmological horizon, one soon 
realises that the asymptotic boundary condition, as proposed by 
Hawking and Page, is no longer applicable. All the solutions to the 
WDW equation turn out to exhibit an oscillatory behaviour when the 
scale factor of the three-geometry becomes greater than the horizon 
size. In fact, the presence of a cosmological horizon in a Lorentzian
spacetime implies that its Euclidean counterpart is compact in 
Euclidean time and, as a consequence, there does not exist an 
asymptotically large Euclidean region around the wormhole end.

In this work, we will propose a generalisation of the asymptotic
wormhole boundary condition for spacetimes with a large cosmological
horizon. According to this proposal, quantum wormholes can be 
represented by wave functions that have an exponentially damped 
WKB behaviour in the region of large three-geometries well inside the
horizon and include only outgoing gravitational modes when the 
three-geometry is asymptotically large. Note, on the other hand, that
the condition that the wormhole wave functions are regular when the 
three-geometry degenerates needs in principle no modification, 
because the presence of a cosmological horizon affects only the large
scale behaviour.

We will particularise our discussion to the simplest gravitational
system that presents a cosmological horizon, namely de Sitter
spacetime. Together with the asymptotically flat and anti-de Sitter
cases (which have already been considered in the literature 
\cite{var,nos2}), our analysis exhausts the study of wormholes 
in maximally symmetric spacetimes. We will see that our generalised 
wormhole boundary condition requires the wormhole throat to 
be much smaller than the existing horizon. Otherwise, the baby 
universe fluctuation could not be distinguished from the background 
spacetime. In this sense, we will understand that the expression 
``quantum wormhole'' refers only to tunnelling processes that 
occur in regions well inside the cosmological horizon. In 
addition, we will assume that the cosmological horizon is large. 

The local effective interactions produced by wormholes have been 
explicitly computed for asymptotically flat \cite{var} and 
asymptotically anti-de Sitter wormholes \cite{nos2} with a variety 
of matter field contents. Using our generalised wormhole boundary 
condition, we will calculate the effective interactions induced by 
wormholes in de Sitter spacetime. We will show that the existence of
a cosmological horizon modifies these interactions with respect to 
the flat case by introducing terms that are proportional to even 
powers of the inverse horizon size.

In section \ref{wf}, we generalise the asymptotic wormhole 
boundary condition and obtain the de Sitter wormhole wave functions.
Sec. \ref{ei} deals with the effective interactions induced by 
these wormholes in low-energy field theory. In both sections,
we work with a conformal scalar field as the matter content.
Finally, we discuss our results and their generalisation
to other matter fields in Sec. 4.

\section{De Sitter wave functions}
\setcounter{equation}{0}
\label{wf}

Let us analyse quantum mechanically a gravitational system with 
positive cosmological constant and a conformally coupled scalar 
field. In the following, $a$ and $\chi_1$ denote the scale factor of
the sections of constant time and the homogeneous mode of the 
conformal scalar field on these sections, respectively. These 
configuration variables will be treated exactly. We will also 
consider the deviations from the homogeneous and isotropic 
configuration described by $a$ and $\chi_1$, but only up to first 
order of perturbation theory. These deviations will be described by 
the coefficients of the expansion of the conformal field in 
hyperspherical harmonics on the three-sphere (in this process, 
gravitational waves are neglected and the gravitational harmonics
are gauged away \cite{nos2, lifs}). Explicitly, we decompose 
the scalar field as\footnote{
>From now on, we will use a rescaled cosmological
constant, $\lambda=\frac{\Lambda}{3}$, and 
set $\frac{2 l_p^2}{3\pi}=1$.}
\begin{equation}
\phi=\sqrt{\frac{1}{2 \pi^2}} a^{-1}
\sum_{n,\sigma_n}\chi_{n\sigma_n} Q^{n\sigma_n},
\label{dsc}
\end{equation}
where $Q^{n\sigma_n}$ are the scalar harmonics, eigenfunctions of 
the Laplace-Beltrami operator in the three-sphere with eigenvalues 
$-(n^2-1)$, the index $\sigma_n$ runs over a basis of the 
corresponding degenerate eigenspace \cite{nos2, lifs},
and the coefficients $\chi_{n\sigma_n}$ depend only on the time
coordinate. In conformal time, the action for the system can then be
written
\begin{eqnarray}
I&=&\int d\eta (\pi_a a'+\pi_{n}\chi_{n}'-NH),  \nonumber \\
H&=&\frac{1}{2}
\left(-\pi_a^2+a^2-\lambda a^4+ \sum_n(\pi^2_{n}-n^2 \chi_{n}^2)
\right).
\label{action}
\end{eqnarray}
Here, $\pi_a$ and $\pi_{n}$ are the momenta canonically 
conjugate to $a$ and $\chi_{n}$, $N$ is the lapse function, and the 
prime denotes derivative with respect to $\eta$. From 
Eq. (\ref{action}), we obtain the WDW equation 
\begin{equation}
\left[-\frac{\partial^2}{\partial a^2}+a^2-\lambda a^4-
\sum_{n,\sigma_n} \left( -\frac{\partial^2}{\partial
\chi_{n\sigma_n}^2} +n^2\chi_{n \sigma_n}^2-\frac{1}{2}
\right)\right] \Psi(a,\chi_{n\sigma_n})=0,
\label{eqfsc}
\end{equation}
in which we have chosen an operator ordering that removes the 
ground-state energy of each of the harmonic oscillators. We can solve
this equation by separation of variables. By imposing standard 
boundary conditions for the quantum harmonic oscillators,
and restricting all considerations to rotationally invariant states 
of the scalar field \cite{nos2,mon}
(i.e., states that depend on the inhomogeneous configuration 
variables only through the rotationally invariant combinations   
$\chi_n^2=\sum_{\sigma_n}\chi_{n\sigma_n}^2 $), we arrive at 
wave functions of the form 
\begin{equation}
\Psi_{N_1,\cdots,N_n,\cdots } (a,\chi_n)= \psi_E(a)
{\cal H}_{N_1}(\chi_1)e^{-\frac{1}{2} \chi_1^2}
\prod_{n>1} {\cal L}_{N_n}^{(n^2-3)/2}(n\chi_n^2)
e^{-\frac{1}{2}n \chi_n^2}, 
\label{wdwsc}
\end{equation} 
where $E=N_1+\sum_{n>1} 2nN_n$ is a sum of harmonic oscillator
energies, ${\cal H}_{N_1}$ is the Hermite polynomial of degree $N_1$,
and ${\cal L}_{N_n}^{(n^2-3)/2}$ is the generalised Laguerre 
polynomial of degree $N_n$ \cite{abra}. The gravitational 
part of the wave function, $\psi_E(a)$, must satisfy the equation
\begin{equation}
\left[-\frac{\partial^2}{\partial a^2}+a^2- \lambda a^4
-2E\right]\psi_E(a)=0.
\label{eqscale}
\end{equation}

If we now imposed the usual wormhole boundary condition that requires
the wave function to be damped when $a\rightarrow +\infty$, we would 
obtain $\psi_E=0$. Actually, every non-vanishing solution of Eq.
(\ref{eqscale}) is asymptotically oscillatory.
 
The potential term $(a^2- \lambda a^4-2E)$ in Eq. (\ref{eqscale})
has two positive roots when $8E\lambda<1$, namely,
$a_{\pm}=(2 \lambda )^{-1/2} [1 \pm (1-8E\lambda)^{1/2}]^{1/2}$. 
These turning points divide the sector of positive scale factors 
into three regions:
two oscillatory, Lorentzian domains, $0<a<a_{-}$ and $a>a_{+}$, and 
an exponential, Euclidean domain, $a_{-}<a<a_{+}$ (for 
$8E\lambda\geq 1$ we have an oscillatory behaviour for all $a>0$). 
When $8E\lambda\ll1$, we can perform a WKB analysis in the 
exponential domain, far from the turning points.
This analysis reveals that there are two possible behaviours for the
wave function in this region, namely, the leading term in the WKB 
approximation can either increase or decrease exponentially for 
increasing $a$. For $a_{-}<a<a_{+}$, the leading-order WKB 
approximation is
\begin{equation}
\psi_E(a)\simeq \sum_{\delta=\pm 1} A_{\delta}\,
(a^2-\lambda a^4-2E)^{-1/4}\exp{\left(\delta\int_{a_{-}}^a 
da' (a^2-\lambda a^4-2E)^{1/2}\right)},
\label{wbk}
\end{equation}
where $A_{\pm}$ are constants.

When $8E\lambda\ll 1$ and $\lambda\ll 1$, this approximation is 
valid \cite{wbkc} at least for scale factors near the bottom of the 
potential, $a_m=1/\sqrt{4\lambda}$. In particular, note 
that (for fixed $E$) the approximation is valid in a region that
overlaps with that of large scale factors if $\lambda$ is 
sufficiently small. Demanding that the leading-order WKB 
approximation is exponentially damped in the region of large scale 
factors well inside the horizon does not totally remove a 
subdominant contribution from the increasing exponential ($\delta=1$)
in Eq. (\ref{wbk}). In the flat case ($\lambda=0$), the increasing 
exponential would dominate the wave function when $a$ becomes 
unrestrictedly large, even if $A_{+}$ is considerably small. Hence, 
the condition that $\psi_E(a)$ is exponentially damped for
asymptotically large scale factors actually implies $A_{+}=0$
if the cosmological constant vanishes. In our case, however, the 
requirement of exponentially damped behaviour only implies
that the quotient $|A_{+}|/|A_{-}|$ (which can depend on
$\lambda$, $E$ and Planck length) has to be small enough 
to suppress the contribution of the increasing exponential in
$\psi_E(a)$ when the scale factor approaches the horizon size.

On the other hand, the de Sitter wave functions display an 
oscillatory behaviour for scale factors larger than the horizon 
size, $a>a_{+}\simeq1/\sqrt{\lambda}$, owing to the presence 
of the cosmological term in the potential. In this region, the WKB 
approximation gives
\begin{equation}
\psi_E(a)\simeq\sum_{\delta=\pm 1} A'_{\delta}\,
(2E-a^2+\lambda a^4)^{-1/4}\exp{\left(\delta i\int_{a_{-}}^a 
da' (2E-a^2+\lambda a^4)^{1/2}\right)},
\label{osc}
\end{equation}
which is always valid for sufficiently large scale factors.
When $A'_{+}=0$, the gravitational wave function $\psi_E(a)$
is completely characterised in the region beyond the horizon
by the fact that it represents a purely outgoing wave. 
Moreover, integrating that solution backwards in $a$, one obtains a 
unique wave function among those which exhibit an exponentially
damped WKB behaviour in the Euclidean region. 

Based on the above discussion, we now introduce the following
proposal.
For sufficiently small cosmological constant, one can interpret as 
quantum wormholes in de Sitter spacetime the wave functions 
(\ref{wdwsc}) whose gravitational part: {\it i}) admits a WKB 
approximation, which is exponentially damped, in the interval of 
large scale factors well inside the Euclidean domain, and {\it ii}) 
corresponds to an outgoing mode in the region beyond the horizon.
Any linear combination of such wave functions represents also a 
quantum wormhole state. Our proposal restricts the existence
of wormhole wave functions to the sector of matter energies
with $8E\lambda\ll1$, since it is only then that an Euclidean region
with the required properties exists. As we have seen, this 
proposal picks out a unique wave function for each value of $E$. 
Finally, notice that the condition that the wave function includes
only outgoing gravitational modes for very large scale factors 
is similar in spirit to the tunnelling proposal of Vilenkin
\cite{vil}, although in our case the tunnelling to the large
Lorentzian region does not occur from ``nothing'', but from
another Lorentzian domain, namely that with small scale factors.

A motivation for this proposal comes from the following 
considerations. For sufficiently small $\lambda$ , let us choose a 
matter energy such that $8E\lambda\ll1$. The turning point $a_{-}$, 
which corresponds to the wormhole throat, is then approximately
equal to $\sqrt{2E}$, which is the throat size of an asymptotically 
flat wormhole with the same matter energy \cite{hall}. In the 
interval $(0, a_{-})$ the de Sitter wormhole wave function has an 
oscillatory behaviour. This behaviour is similar to that displayed 
by a flat wormhole \cite{pag} and describes a Lorentzian closed 
Friedmann-Robertson-Walker spacetime, i.e. a baby universe. 
Furthermore, in the region 
$\sqrt{2E}\ll a\ll a_{+}\simeq1/\sqrt{\lambda}$, 
where the potential is dominated by the term $a^2$, one can parallel
the line of reasoning discussed by Hawking and Page in the flat case 
to conclude that the main contribution to the wormhole wave function
in the saddle point approximation must be given by the exponential 
of the surface term $-\frac{1}{2}\int\sqrt{h}|K|d^3x$, evaluated on 
the section of constant time with scale factor $a$. Here, $K$ is the
trace of the extrinsic curvature. It is then easy to see that one 
arrives precisely at the exponentially damped WKB behaviour that we 
have proposed for the wave function $\psi_E(a)$ in the region of 
scale factors under consideration. In order to select a wave 
function among those that possess this exponentially damped 
behaviour, one needs to generalise the arguments given by Hawking 
and Page to the region of scale factors beyond the horizon. In this 
region, the saddle points are Lorentzian and describe asymptotically
de Sitter geometries that either expand or contract from a scale 
factor $a$ larger than the horizon size. When the dominant 
saddle points are expanding (contracting) geometries, the wave 
function includes only outgoing (ingoing) gravitational modes 
beyond the horizon. In the flat case, the condition that 
the wormhole wave functions are asymptotically damped reflects 
the fact that the tunnelling occurs from the baby universe 
to the region of asymptotically large three-geometries, but not in 
the opposite direction. It then seems natural to generalise this 
condition to the de Sitter case by imposing that, beyond the 
cosmological horizon, there are only outgoing gravitational modes, 
so that the wave function describes in fact a tunnelling from small 
to asymptotically large scale factors. In this sense, our proposal 
provides a natural generalisation of the asymptotic wormhole 
boundary condition once it is assumed that the size of the 
cosmological horizon ($1/\sqrt{\lambda}$) is sufficiently large. 
The de Sitter wormhole wave functions selected by this proposal 
have a behaviour that is similar to that of the asymptotically 
flat wormholes both in the baby universe sector and in the region 
of large scale factors much smaller than the horizon size.
  
Since $E\lambda$ is the square of the rate between the wormhole 
throat (of order $\sqrt {E}$) and the horizon size of de Sitter 
space ($1/\sqrt{\lambda}$), the condition $8E\lambda\ll 1$, that 
we have used in our discussion, allows us to regain the picture of 
wormhole connections in a background spacetime. This picture would
break down for quantum states with $8E\lambda \gsim 1$, because 
these states describe quantum fluctuations whose characteristic size 
is of the order of or greater than the cosmological horizon of the 
de Sitter background. Hence, we will not regard the states 
with $8E\lambda\gsim 1$ as quantum wormholes.

Finally, we will see in the next section that the wave functions 
(\ref{wdwsc}) with the above choice for $\psi_E(a)$ can be 
interpreted as de Sitter wormholes with a definite number of 
particles in the Euclidean vacuum for de Sitter space 
\cite{tag,allen3}, that is, the vacuum which is conformally related 
to the natural vacuum for flat spacetimes \cite{nos2}.

\section{Effective interactions.}
\setcounter{equation}{0}  
\label{ei}

Following a procedure developed by Hawking \cite{haw}, and used
in Ref. \cite{nos2} for the anti-de Sitter case, we will now 
deduce the explicit form of the effective interactions 
produced by wormholes in de Sitter spacetime. In order to do this, 
we must first calculate the matrix elements of products of matter 
fields between a vacuum in de Sitter spacetime and an arbitrary 
wormhole state, $|\Psi_{\alpha}\rangle$,
\begin{equation}
\langle \Psi _\alpha |\phi (x_1)\phi (x_2)| 0 \rangle.   
\label{scme}
\end{equation}
Here, we have considered only two matter fields for symplicity. 
In this section, we will choose the vacuum $| 0 \rangle$ to be the 
Euclidean vacuum. In Sec. 4, we will comment on the consequences 
of different choices of vacuum. Since our aim is
to determine the effects of wormholes in scales greater that the 
wormhole scale,  $x_1$ and $x_2$ represent points in regions of the 
Euclidean de Sitter spacetime far from the tiny wormhole end. For 
scale factors $a$ in the Euclidean region, the state
$\phi (x_1)\phi (x_2)| 0 \rangle$ is then given, in a 
$(a, \chi)$-representation, by a Euclidean path integral over 
geometries and matter field configurations with initial values $a$ 
and $\chi$ for the scale factor and the matter field, respectively, 
and compatible with the condition $E_{\chi}=E_a=0$ at an arbitrary 
final time. Here, 
$E_{\chi}=\frac{1}{2}\sum_n (n^2\chi_n^2-\pi^2_{\chi_n})$ is
the energy of the matter field, and 
$E_a=\frac{1}{2}(a^2-\lambda a^4- \pi_a^2)$ is the energy associated
with the scale factor. Fixing the variables $E_{\chi}$ and $E_a$ in 
this way at a final time $\tau_f$ selects the Euclidean vacuum
for the matter field (the classical solutions for the scalar field 
with $E_{\chi}=0$ correspond to the Euclidean mode decomposition 
of the field) and makes the action invariant under time 
reparametrizations that coincide with the identity at the initial
time, but are arbitrary at $\tau_f$ \cite{nos1}. Therefore, the 
particular value chosen for the final time becomes irrelevant. 

An estimate of the path integral can be obtained 
by means of a saddle point approximation. As far as the low-energy 
regime is concerned, the geometrical saddle point can be taken to
be pure de Sitter space outside a subtracted sphere of radius
$a$ that surrounds the wormhole insertion. Then, the saddle point 
solution for the conformal scalar field must satisfy the equation 
$(\Box -2\lambda )\phi=0$, where $\Box$ is the Laplacian for the de 
Sitter four-sphere. If $\phi(x')$ is a saddle point solution, 
$\phi_f(x')$, the transform of $\phi$ under an element $f^{-1}$ of 
the group of isometries $SO(5)$, is also a solution. As a 
consequence, one has to average over the group $SO(5)$. Recalling 
that the four-sphere $S^4$ (i.e., Euclidean de Sitter spacetime)
and the coset space $SO(5)/SO(4)$ are isomorphic, 
the integral over $SO(5)$ can be performed as follows
\begin{equation}
\int_{SO(5)} df F(f) = \int_{S^4} d^4x \sqrt{g(x)} 
\int_{SO(4)} dh F(x h),
\end{equation}
$h$ being a generic element of the isotropy group $SO(4)$ and $g$
the determinant of the metric on the four-sphere.
This integral can be interpreted as an average over the positions
and orientations in which a wormhole end can be connected. 

In Ref. \cite{nos2} it was shown, by treating each mode separately, 
that the different saddle points can be expressed in terms of the 
propagator of the matter field as
\begin{equation} 
\phi_{xh}^{n}(x')={\cal M}_h^n \cdot \Theta_n G(x',x),
\label{sp}
\end{equation} 
where $G$ is the propagator associated with the Euclidean vacuum,
and ${\cal M}_h^n$ is a constant tensor of range $n-1$ that is 
completely symmetric and vanishes under contractions of any pair of 
indices. This tensor contains all the dependence on the rotation
group, as well as on $a$ and $\chi_n$ (i.e., the values of the scale 
factor and the matter field coefficients on the sphere that has been 
subtracted to de Sitter space). The function $\Theta_n G$ is 
constructed by completely symmetrising a product of $n-1$ covariant 
derivatives acting on the $x$-de\-pendence of the propagator, 
$\nabla^{\mu_1} \cdots \nabla^{\mu_{n-1}}G$, and subtracting all its
traces \cite{nos2}. Finally, the dot in (\ref{sp}) denotes the scalar
product in the linear space of tensors with the explained symmetries. 
Taking into account that ${\cal M}_h^n=R(h){\cal M}^n$, 
where $R(h)$ is the appropriate irreducible representation of
the rotation group, which satisfies  
\begin{equation}
\int_{SO(4)}dh R(h) \otimes R(h)= 1 \otimes 1
\end{equation}
(with $\otimes$ the tensor product), the average over orientations 
of the product of the two matter field solutions leads to
\begin{equation}
\int_{SO(4)}dh \phi_{xh}^{n}(x_1)\phi_{xh}^{n}(x_2)=
({\cal M}^n \cdot {\cal M}^n) \left(\Theta_n G(x_1,x) \cdot 
\Theta_n G(x_2,x) \right).
\end{equation}
Hence, the complete expression for the quantum state
$\phi (x_1)\phi (x_2)| 0 \rangle$ has the form
\begin{equation}
F(a,\chi_n)
\int_{S^4} d^4x \sqrt{g(x)}(\Theta_n G(x_1,x) \cdot \Theta_n
 G(x_2,x) ).
\label{meexpr}
\end{equation}
In the function $F(a,\chi_n)$, the dependence on $\chi_n$ is of the 
form $(\chi_n)^2e^{-\frac{1}{2} n^2 (\chi_n)^2}$ \cite{nos2}, where 
the factor of $(\chi_n)^2$ comes from the product 
$({\cal M}^n \cdot {\cal M}^n)$ and the exponential factor from 
the evaluation of the action on the classical solution. 

The orthogonality of the Hermite and Laguerre polynomials that 
appear in (\ref{wdwsc}) implies then that the state
$\phi (x_1)\phi (x_2)| 0 \rangle$ has non-vanishing projections
only on the vacuum and either on the $N_n=1$ state for $n>1$ 
or the $N_1=2$ state for the homogeneous case. We can therefore 
interpret $\Psi_{N_n=1}$ with $n>1$ and $\Psi_{N_1=2}$ as 
quantum wormholes that, in the Euclidean vacuum, contain a 
two-particle rotationally invariant state and two homogeneous 
particles, respectively. A similar analysis can be applied as well 
to the three-point function and higher functions, extending the 
above interpretation to any wave function 
$\Psi_{N_1, \cdots, N_n \cdots}$ of the form (\ref{wdwsc})
such that $(N_1+\sum_{n>1} 2nN_n)8\lambda=8E\lambda\ll1$.

One can now deduce the expression of the interaction 
Lagrangian that, via the formula $\langle 0|\phi(x_1)\phi(x_2)\int 
d^4x\sqrt{g(x)}{\cal L}_n^I(\phi(x)) | 0 \rangle$, reproduces the 
matrix element (\ref{scme}) up to a constant factor. 
This Lagrangian must be of the form 
$ {\cal L}_n^I= \Theta_n \phi \cdot \Theta_n \phi$,
as can be seen by making use of Wick's theorem and noting that the
operator $\Theta_n$ is linear \cite{nos2}. 
For the lowest modes, for instance, the 
interaction Lagrangians turn out to be
\begin{equation}
{\cal L}_1^I=\phi^2, \hspace{0.7cm} 
{\cal L}_2^I=\nabla^{\mu} \phi\nabla_{\mu}\phi, \hspace{0.7cm}
{\cal L}_3^I=
\left(\nabla^{\rho}\nabla^{\sigma}\phi
-\frac{1}{2}\lambda
g^{\rho\sigma}\phi\right)
\left(\nabla_{\rho}\nabla_{\sigma}\phi
-\frac{1}{2}\lambda
g_{\rho\sigma}\phi\right).
\end{equation}
It is worth noting that, for $n \geq 3$, the form of the 
interactions differs from that obtained for flat space in terms that
depend on $\lambda$, the square of the inverse horizon size. 
Therefore, the local interactions introduced by wormholes seem to 
depend on the large structure of spacetime. It then might happen 
that the constants of nature could be affected by contributions 
of cosmological origin owing to the existence of quantum wormholes.

\section{Discussion and conclusions.}
\setcounter{equation}{0}
\label{dac}

In this work, we have analysed the possible effects of the 
existence of wormholes in cosmological spacetimes with matter 
content. In this kind of spacetimes, it is necessary to
generalise the standard, asymptotic wormhole boundary condition,
because, when the spacetime possess cosmological horizons, no
asymptotic Euclidean region exists. We have considered in detail
the case of de Sitter spacetime, and extended the line of
reasoning discussed by Hawking and Page for the case of flat 
wormholes. In this way, we have arrived at the following proposal.
In a large de Sitter spacetime (i.e., when the cosmological constant
is sufficiently small), it is possible to interpret as quantum 
wormhole states the wave functions that {\it i}) admit a WKB 
approximation with exponentially damped leading term in the region 
of large scale factors much smaller than the horizon size, and 
{\it ii}) contain only outgoing gravitational modes beyond the 
horizon. Unlike the situation found in the flat and anti-de Sitter 
cases, the existence of a cosmological horizon in de Sitter space
poses an obstruction for the interpretation of a quantum state as a 
wormhole: the interpretation is feasible only in the sector of 
states with small matter energy. This restriction is necessary to 
guarantee that there exists a large Euclidean region in which the 
wave functions can have an exponentially damped behaviour. Without 
this restriction in the matter energy, the entire observable 
universe could be contained in the considered quantum fluctuation, 
so that it would be imposible to distinguish the interior of the 
wormhole from the background universe.

We have analysed the case of de Sitter wormholes with a conformal 
scalar field and discussed the effects of these quantum fluctuations
in low-energy field theory. We have shown that the 
effective interactions produced by these wormholes
differ from those induced by flat wormholes for $n=3$ and higher 
harmonics. These differences are proportional to even powers of the 
inverse horizon size, i.e., to positive powers of the cosmological 
constant.

For other matter fields one would obtain similar results. As long 
as there exist quantum wormholes that admit the interpretation 
of small connections in a background spacetime, one can calculate 
their effective interactions in the following way. 
Given a field with spin $s$, each of the 
hyperspherical harmonics on the three-sphere in which we can 
decompose its true degrees of freedom carries an irreducible 
representation of the group 
$SU(2) \otimes SU(2)$, the universal covering of the isotropy 
group $SO(4)$. This irreducible representation is
of type $(m/2+s,m/2)$ or $(m/2,m/2+s)$, where $m=n-s-1$ is a non 
negative integer, $n$ is the mode of the considered harmonic, and 
the lowest mode is given by $n=s+1$. Each harmonic gives rise to a 
different interaction Lagrangian. The explicit form of these 
Lagrangians would be $\Theta_n \Phi \cdot \Theta_n \Phi$, with 
$\Phi$ representing a matter field of spin $s$.  Finally, the 
operator $\Theta_n$ can be constructed with the help of the 
symmetries that the corresponding hyperspherical harmonic possesses 
when we write it in a Cartesian basis \cite{nos2, lifs}, as it was 
proved in Ref. \cite{nos2} for integer spins.

Let us illustrate this point with two examples. We will first 
consider the case of de Sitter wormholes with a minimally coupled 
massless scalar field $\phi$. In the process of subtracting
all the traces of the completely symmetric product of covariant 
derivatives that act on the propagator (as was explained above for 
the conformal scalar field), one must take into account that
the approximate saddle point equation is now
$g^{\mu \nu} \nabla_{\mu}  \nabla_{\nu}\phi=0$, instead of the 
equation that applies in the conformally coupled case,
$(g^{\mu \nu} \nabla_{\mu}  \nabla_{\nu}-2\lambda)\phi=0$. This 
results in changing the interaction Lagrangians for the third 
($n=3$) and higher modes with respect to those obtained for the 
conformal field. For instance, the Lagrangian for the mode $n=3$ 
takes the form ${\cal L}_I^3=(\nabla_{\mu}\nabla_{\nu}\phi)^2$. 
In contrast with the conformal field case,
the interaction Lagrangians for the minimally coupled scalar field 
in flat and de Sitter spaces differ then just in that partial 
derivatives in flat space are replaced with covariant derivatives in
de Sitter space. Let us discuss now the case of an electromagnetic 
field. In this case, $\Phi$ represents the four-potential $A_{\mu}$. 
The interaction Lagrangian for the lowest mode  
$n=2$ reads $(\Theta_2\Phi)^2=
(\nabla_{\mu}A_{\nu}-\nabla_{\nu}A_{\mu})^2=(F_{\mu\nu})^2$
\cite{nos2}. One can then recursively construct $\Theta_n \Phi$ 
for higher harmonics by symmetrising $\nabla\Theta_{n-1} \Phi$ and 
subtracting all its traces. This subtraction is easily performed by 
noticing that $\Box F_{\mu\nu}=4 \lambda F_{\mu\nu}$, an equation
that follows from $\nabla^{\mu}F_{\mu\nu}=0$. Finally, the tensor
$(\Theta_{n}\Phi)_{\mu_1 \mu_2, \mu_3 \cdots \mu_n}$ must be 
antisymmetric in its first two indices, symmetric with respect to 
all other indices, and vanishing when contracted in any pair of 
indices or when taking a cyclic sum over $\mu_1$, $\mu_2$, and any 
other index. These are the symmetries corresponding to the 
$n$-th mode of the transverse vector harmonics on the 
three-sphere, which are the true degrees of freedom of the 
electromagnetic field \cite{nos2}. On the other hand, when dealing 
with the electromagnetic field (as well as with other fields of 
higher spin), another physical quantity comes into play: the 
helicity. The helicity distinguishes the $(p,q)$ and $(q,p)$ 
irreducible representations. This can be done by introducing 
operators $\Theta_{n\pm}$ that are the self-dual and anti-self-dual 
parts of $\Theta_n$. Nonetheless, the interaction Lagrangians for 
positive and negative helicities turn out to coincide,
because the cross product $\Theta_n \Phi \cdot {^{*}\Theta}_n \Phi$
is a topological invariant (as can be cheked by direct calculation).

In general, we find that the wormhole effective interactions have 
contributions owing to the presence of a cosmological horizon. The 
explicit form of such contributions depends on the specific matter 
content.

To conclude, we will make some comments on our choice of vacuum 
in Eq. (\ref{scme}). Throughout our calculations, the state 
$|0\rangle$ has been chosen as the Euclidean vacuum in de Sitter 
space. In Ref. \cite{nos2}, it was shown that, for anti-de Sitter 
wormholes, the choice of a particular vacuum among the family of 
maximally symmetric vacua is a matter of convenience. Once a vacuum
is chosen, one can always construct a Fock space of quantum 
wormholes labelled by the number of particles that they contain,
as referred to that vacuum. In the de Sitter case, however, the 
existence of the horizon scale implies that the Euclidean vacuum 
plays an special role. The total number of particles associated with 
this vacuum that a de Sitter wormhole can contain is restricted by 
the condition $8N\lambda\ll1$. A quantum state $\Psi_{\bar {N}}$ 
with a definite number of particles $\bar {N}$ in another vacuum has 
non-vanishing projections in states with $8N\lambda\geq 1$ regardless 
of the value of $\bar{N}$, so that it cannot be interpreted as a
quantum wormhole. Therefore, the Euclidean vacuum turns out to
be special in the sense that only observers associated with it 
can interpret certain wormhole states as containing a definite 
number of particles. 
 
\section*{Acknowledgments}

The author is very grateful to G. A. Mena Marug\'an and L. J. Garay
for useful suggestions and discussions and for a critical reading
of the manuscript. He also would like to thank P. F. Gonz\'alez-D\'{\i}az
for many enlightening comments on the subject of this paper.
The author was supported by a Spanish Ministry of Education and
Culture (MEC) grant.

\end{document}